\documentstyle[12pt]{article}

\renewcommand{\theequation}{\thesection.\arabic{equation}}
\newcommand{\be}{\begin{equation}}   \newcommand{\ee}{\end{equation}}
\newcommand{\bear}{\begin{eqnarray}}
\newcommand{\eear}{\end{eqnarray}}
\newcommand{\ba}{\begin{array}}      \newcommand{\ea}{\end{array}}
\newcommand{\lae}{\begin{array}{c}\,\sim\vspace{-21pt}\\< \end{array}}
\newcommand{\gae}{\begin{array}{c}\,\sim\vspace{-21pt}\\> \end{array}}
\newcommand{\up}{\mbox{$P$}}
\newcommand{\down}{\mbox{$N$}}
\newcommand{\inl}{{\scriptscriptstyle L}}
\newcommand{\inr}{{\scriptscriptstyle R}}
\renewcommand{\theequation}{\thesection.\arabic{equation}}

\newcommand{\drawsquare}[2]{\hbox{%
\rule{#2pt}{#1pt}\hskip-#2pt
\rule{#1pt}{#2pt}\hskip-#1pt
\rule[#1pt]{#1pt}{#2pt}}\rule[#1pt]{#2pt}{#2pt}\hskip-#2pt
\rule{#2pt}{#1pt}}
\newcommand{\Yfund}{\raisebox{-.5pt}{\drawsquare{6.5}{0.4}}}
\def\vbr{$\vphantom{\sqrt{F_e^i}}$}

\begin{document}

\pagestyle{empty}
\begin{titlepage}
\def\thepage {}        

\title{\bf  Top-Bottom Splitting in Technicolor \\ [2mm] with Composite Scalars
\\ [1cm]}

\author{\bf Bogdan A.~Dobrescu$^1$ and  Elizabeth H.~Simmons$^{2}$ \\
\\
{\small {\it $^1$Fermi National Accelerator Laboratory}}\\
{\small {\it Batavia, IL 60510, USA \thanks{e-mail
  addresses: bdob@fnal.gov, simmons@bu.edu} }}\\
\\
{\small {\it $^2$Department of Physics}} \\
{\small {\it Boston University}}\\
{\small {\it Boston, MA 02215}} \\ }

\date{ }

\maketitle

   \vspace*{-13.cm}
\noindent
\makebox[12.3cm][l]{FERMI--PUB--98/225-T} hep-ph/9807469 \\ [1mm]
\makebox[12.3cm][l]{BUHEP-98-16}  July 22, 1998\\

   \vspace*{12.cm}

\baselineskip=18pt

\begin{abstract}

   {\normalsize
We present a model of dynamical electroweak symmetry breaking
in which the splitting between the top and bottom quark masses arises
naturally. The $W$ and $Z$ masses are produced by a minimal technicolor 
sector, the top quark mass is given by the exchange of a weak-doublet
technicolored scalar, and the other quark and lepton masses are induced 
by the exchange of a weak-doublet technicolor-singlet scalar.
We show that, in the presence of the latter scalar, the vacuum alignment 
is correct even in the case of $SU(2)$ technicolor. The fit of this model to 
the electroweak data gives an acceptable agreement ($\chi^2 = 28$, for 20 
degrees of freedom).
The mass hierarchy between the standard fermions other than top can also be
explained in terms of the hierarchy of squared-masses of some 
additional scalars. 
We discuss various possibilities for the compositness of the scalars 
introduced here. 
   }
   
\end{abstract}

\vfill
\end{titlepage}

\baselineskip=18pt
\pagestyle{plain}
\setcounter{page}{1}

\section{Introduction}
\setcounter{equation}{0}

While the dynamics responsible for the generation of mass remains obscure,
there are a few known theoretical possibilities that explain certain
relationships between the masses of the observed particles.  The success of
the standard model in fitting the experimental results may appear to favor
models that include a Higgs boson in the low energy effective theory, such as
supersymmetric standard models or top condensation models
\cite{tmode,tmoderec}.  Yet the current precision of the electroweek
measurements does not actually distinguish between the standard model and
certain models that do not have a decoupling limit.  The latter theories use
technicolor to give the $W$ and $Z$ masses, and additional fields
to communicate electroweak symmetry breaking to the quarks and leptons. If
these fields are heavy gauge bosons, as in extended technicolor \cite{etc},
then one is led to consider complicated dynamics \cite{topassist,ncETC}.

On the other hand, if the additional fields are scalars, one has the
flexibility to generate the observed masses without immediate dynamical
assumptions.  For example, technicolor models with weak-doublet
technicolor-singlet scalars \cite{tcscal, carsim,massless,phen} have been
found to have phenomenology consistent with experiment.  Alternatively,
technicolor models that include weak-singlet technicolored scalars
\cite{kagan1, kagan2, susytc,negs} give a natural explanation for the mass
hierarchy between the fermion generations.  The existence of scalars much
lighter than the Planck scale does require some further explanation.  Their
masses can be protected by supersymmetry \cite{tcsd, bostc,kagan1,susytc}, or
they can be bound states arising within a high energy theory\cite{strongETC}.
It is also conceivable that the fundamental scale where quantum gravity
becomes strong is not $10^{19}$ GeV, but rather some TeV
scale \cite{lykken}.

In this paper we show that a technicolor model that includes weak-{\it
  doublet} technicolored scalars explains not only the inter-generational
fermion mass hierarchy, but also the intra-generational mass hierarchies, in
terms of relationships among the squared masses of different scalars. For
example the top-bottom splitting arises naturally in such models because
hypercharge prevents the techniscalar responsible for the top quark mass from
inducing a bottom quark mass.

We start by constructing the low-energy effective theory that gives rise to
the $W$, $Z$ and $t$ masses, without specifying a dynamical origin for the
scalars.  In section 3 we explore the elecroweak phenomenology of the
low-energy effective theory. Next, we discuss possible mechanisms for
generating the masses of the other quarks and leptons.  Dynamics that could
create the scalar bound states in our models are addressed in section 5.  We
present conclusions in section 6.  In Appendix A we show that $SU(2)$
technicolor breaks the electroweak symmetry correctly in the presence of the
scalar used to give mass to the light fermions. In Appendix B we present the
fit to the electroweak data.

\section{Technicolor and the Top Mass} 
\setcounter{equation}{0}

Our model includes the standard model gauge and fermion sectors together
with a minimal technicolor sector intended to break the electroweak
symmetry dynamically.  The latter consists of an asymptotically free 
$SU(N_{\rm TC})$ gauge group, which becomes strong at a 
scale of order 1 TeV, and one doublet of technfermions 
which transform under the $SU(N_{\rm TC}) \times SU(3)_{\rm C}
\times SU(2)_{\rm W} \times U(1)_{\rm Y}$ gauge group as:
\be
\Psi_{\!\!\inl} = 
\left(\!\!\ba{c}\up_{\inl} \\ \down_{\inl}\ea\!\!\right) 
  \, : \; (N_{\rm TC}, 1, 2)_0 \; , \hspace{.5cm}
\up_{\inr} \, : \;  (N_{\rm TC}, 1, 1)_{+1}  \; , \hspace{.5cm}
\down_{\inr} \, : \;  (N_{\rm TC}, 1, 1)_{-1} \; .
\label{e1}
\ee
The dynamics of the technicolor interactions is taken from QCD:
the $SU(2)_L \times SU(2)_R$ chiral symmetry of the technifermions 
is spontaneously broken by the condensates
\be
\langle \overline{\up} \up \rangle \approx 
\langle \overline{\down} \down \rangle \approx 
4 \pi f^3 \left(\frac{3}{N_{\rm TC}} \right)^{\! 1/2} ~,
\label{cond}
\ee
where $f$, the technipion decay constant, is the analog of $f_\pi$ in
QCD.  Since the $SU(2)_{\rm W} \times U(1)_{\rm Y}$ group is embedded in
the chiral symmetry, the technifermion condensates break the electroweak
symmetry. If minimal technicolor is the only source of electroweak
symmetry breaking, then the observed $W$ and $Z$ masses require $f = v$,
where $v \approx 246$ GeV is the electroweak scale.

The only constraints on this electroweak symmetry breaking mechanism
come from the oblique radiative correction parameter $S$, which measures
the momentum-dependent mixing of the neutral electroweak gauge bosons.
The technifermion contribution to $S$ can be estimated by using the QCD
data \cite{peskin,sdef}:
\be
S \approx 0.1 \; N_{\rm TC} ~.
\label{spar}
\ee
A fit to the electroweak data (using the standard model with a Higgs mass of
300 GeV as a reference) yields \cite{negs} a 1$\sigma$ ellipse in the $S-T$
plane whose projection on the $S$-axis is $S = - 0.09 \pm 0.34$.  Thus, $S$
in the minimal technicolor model is smaller than the 2$\sigma$ upper bound
provided $N_{\rm TC} < 6$.  The cancelation of the Witten anomaly for
$SU(2)_W$ requires $N_{\rm TC}$ to be even.  If the only interactions, in
addition to technicolor, experienced by the technifermions were the
electroweak interactions, then the value $N_{\rm TC} = 2$ would be ruled out
because the most attractive channel for condensation, $\langle \up_L\down^c +
\down_L\up^c \rangle$, breaks the electroweak group completely \cite{align}.
However, the generation of quark and lepton masses requires additional
interactions of the technifermions, which may easily tilt the condensate in
the correct direction. For example, in section 4.1, we introduce a
weak-doublet scalar to communicate electroweak symmetry breaking to the light
fermions.  As shown in Appendix A, the scalar's interactions with the
technifermions would have a sufficiently large effect on the technifermion
condensate to make the case $N_{\rm TC} = 2$ viable.  Therefore, in what
follows we adopt the values 
\be 
N_{\rm TC} = 2, \, 4 ~.  
\ee

In order to generate the large top quark mass, we have to specify some new
physics at a scale of order 1 TeV that allows the minimal technicolor sector 
discussed so far to couple to the top. A particularly attractive alternative 
is to introduce a scalar multiplet, $\chi^t$, which transforms
under the $SU(N)_{\rm TC} \times SU(3)_{\rm C} \times SU(2)_{\rm W}
\times U(1)_{\rm Y}$ gauge group as:
\be
(\overline{N_{\rm TC}}, 3, 2)_{4/3} ~.
\ee
The most general Yukawa interactions are contained in 
\be 
{\cal L}_{t} = C_q^j\, \overline{q_{\inl}^{\prime j}}\, \down_{\inr}\,
\chi^t \, +\, C_u^j\, \overline \Psi_{\!\!\inl}\, u_{\inr}^{\prime j}\, i
\sigma_2\,\chi^{t \dagger} 
\, +\, {\rm h.c.} ~,
\label{yukawas}
\ee
where $j \in {1,2,3}$ counts the generations, $\sigma_i$ are the Pauli
matrices; $C_q^j$, $C_u^j$ are Yukawa couplings,
$u_{\inr}^{\prime j}$ are the right-handed up-type quarks, and
$q_{\inl}^{\prime j} \equiv (u_{\inl}^{\prime j}, d_{\inl}^{\prime
  j})^\top$ are the left-handed quarks, defined in an arbitrary
eigenstate.  At first glance, it appears that all three generations of
right-handed up-type and left-handed quarks couple to the $\chi^t$
scalar. However, these couplings are linear in the quark fields (unlike
the bilinear quark couplings to the Higgs doublet in the standard
model), and therefore only one linear combination of the three generations
couples to $\chi^t$.  Because this is the combination that becomes heavy,
it is identified by convention with the third generation in the weak
eigenstate.  Therefore, the Lagrangian  in eq.~(\ref{yukawas}) is equivalent to
\be 
{\cal L}_{t} = C_q\, \overline{q^3_{\inl}}\, \down_{\inr}\, \chi^t 
\, +\, C_t\, \overline \Psi_{\!\!\inl}\, t_{\inr}\, i
\sigma_2\,\chi^{t \dagger} 
\, +\, {\rm h.c.} ~,
\label{tyuk}
\ee
where $q_{\inl}^3  \equiv (t_{\inl}, b_{\inl})^\top$ is the left-handed 
weak eigenstate $t-b$ quark doublet, and
the Yukawa coupling constants, $C_q$ and $C_t$, are defined to be positive.

If the mass of the $\chi^t$ scalar is much larger than $f$, then at
energies of the order of the electroweak scale the effects of
$\chi^t$ exchange are well-described by the following four-fermion
operators:
\bear
{\cal L}_{\rm 4F} = & - & \frac{1}{2 M_{\chi^t}^2}
\left\{ C_q^2 \left(\overline{\down}_{\!\!\inr} \gamma^\mu \down_{\!\!\inr} 
\right)
        \left(\overline q^3_{\inl} \gamma_\mu q^3_{\inl} \right)
+ C_t^2 \left(\overline{\Psi}_{\inl} \gamma^{\mu} \Psi_{\inl} \right)
        \left(\overline{t}_{\inr}\gamma_{\mu} t_{\inr} \right)\right.
\nonumber \\ [3mm] 
& + & \left. \left[ C_qC_t \left(\overline{\Psi}_{\inl}\down_{\!\!\inr} \right)
        \left(\overline{t}_{\inr} q^3_{\inl} \right)
+ \frac{C_qC_t}{4} 
        \left(\overline{\Psi}_{\inl} \sigma^{\mu\nu} \down_{\!\!\inr} \right)
        \left(\overline{t}_{\inr} \sigma_{\mu\nu} q^3_{\inl} \right) \right]
 + {\rm h.c.} \right\} ~,
\label{operator}
\eear
Upon technifermion condensation, the third operator in eq.~(\ref{operator})
induces a top mass
\be
m_t \approx \frac{C_qC_t}{M_{\chi^t}^2}\pi f^3 
\left(\frac{3}{N_{\rm TC}} \right)^{\! 1/2} ~,
\label{mtop}
\ee
Using $m_t \approx 175$ GeV and $f \approx 246$ GeV we get
\be
M_{\chi^t} \approx 570 \; {\rm GeV} \sqrt{C_qC_t} \left(\frac{2}{N_{\rm TC}} 
\right)^{\! 1/4} ~,
\label{mchival}
\ee
which shows that the assumption $M_{\chi^t} \gg f$ is valid only if
the Yukawa coupling constants are rather large. This situation seems
plausible if $\chi^t$ is a bound state, but in this case loop
corrections to the operators in eq.~(\ref{operator}) might need to be included
in the low energy theory.  Alternately, for $C_q$ and $C_t$ of order one
or smaller, $M_{\chi^t}$ is not much larger than $f$ and the
technicolor dynamics might be modified by the existence of $\chi^t$.
With these limitations in mind, we will assume that the effects of
$\chi^t$ are described sufficiently well by the operators in 
eq.~(\ref{operator}).

It is remarkable that the hypercharge of $\chi^t$ allows it to couple to
$t_R$ but not to $b_R$.  As a consequence, with the field content discussed
so far, the only standard fermion that becomes massive is the top quark.
Provided that the other quark and lepton masses are produced by physics above
the technicolor scale, it will be natural for the top quark to be the
heaviest fermion. This situation is in contrast with the case of a weak-{\it
  singlet} technicolored scalar, which can couple to both $t_R$ and $b_R$
\cite{kagan2,susytc,negs} and needs the top-bottom mass ratio to be provided
by a ratio of Yukawa couplings.  Note that the models with
weak-singlet techni-scalars naturally explain the small CKM elements
associated with the third-generation quarks, because the $t_L$ and $b_L$ mass
eigenstates are automatically aligned.

\section{Electroweak Observables and $\chi^t$}
\setcounter{equation}{0}

As mentioned earlier, if the mass of the $\chi^t$ scalar is much larger 
than $f$, at energies below the weak scale the effects of $\chi^t$ 
exchange are captured by the four-fermion operators in 
eq.~ (\ref{operator}).  One consequence of those four-fermion operators 
is the generation of a large mass for the top quark.  Another, 
as we shall now discuss, is a significant contribution to the 
couplings of the $SU(2)_W \times U(1)_Y$
gauge bosons, $W^\mu_i \, (i = 1,2,3)$ and $B^\mu$, to the $t$ and
$b$ quarks.  This causes both direct and oblique corrections to 
electroweak observables.

Below the technicolor scale, only techni-pion dynamics has an impact
on the electroweak observables, and these effects can be evaluated using
an effective Lagrangian approach.  Recalling that $f \approx v$ in our
minimal one-doublet technicolor sector we find:
\bear
\overline \Psi_{\!\!\inl} \gamma^\mu \Psi_{\!\!\inl}         
& = & i \frac{v^2}{2}
{\rm Tr} \left(\Sigma^{\dagger} D^{\mu} \, \Sigma \right)
\nonumber \\ [2mm]
\overline{\down}_{\inr} \gamma^{\mu} \down_{\inr} & = &  - i \frac{v^2}{2}
{\rm Tr} \left(D^{\mu} \Sigma \frac{-\sigma_3+1}{2} \Sigma^{\dagger} \right)
\\ [2mm]
\overline{\up}_{\inr} \gamma^{\mu} \up_{\inr} & = &  - i \frac{v^2}{2}
{\rm Tr} \left(D^{\mu} \Sigma \frac{\sigma_3+1}{2} \Sigma^{\dagger} \right)
\nonumber
\eear
where $\Sigma$ transforms as $W \Sigma R^{\dagger}$ under
$SU(2)_W \times SU(2)_R$ (where $SU(2)_R$ is the global symmetry which
has $U(1)_Y$ as a subgroup). Note that the last two terms in 
eq.~(\ref{operator})
do not affect the techni-pions to leading order.
The covariant derivative is 
\be
D^{\mu} \Sigma = \partial^{\mu} \Sigma -
i g \frac{\sigma^k}{2} W^{\mu}_k \Sigma + i g^\prime \Sigma \frac{\sigma^3}{2}
B^{\mu}
\ee
which gives
\bear
\overline \Psi_{\!\!\inl} \gamma^\mu \Psi_{\!\!\inl} & = & 0
\nonumber \\ [2mm]
\overline{\down}_{\inr} \gamma^{\mu} \down_{\inr}
& = & \frac{v^2}{4} \left(g W^{\mu}_3 - g^\prime B^\mu \right)
=  \frac{v^2}{4} \frac{g}{c_W} Z^\mu
= - \overline{\up}_{\inr} \gamma^\mu \up_\inr \ \ .
\label{vchg}
\eear

The result is that, in addition to the standard model couplings of the 
$SU(2)_W \times U(1)_Y$ gauge bosons to third generation quarks, the following
coupling is induced by the exchange of the $\chi^t$ scalar:
\be
{\cal L}_{\rm eff}  =  \delta g \frac{g}{c_W} Z^\mu
( \overline q_{\inl} \gamma_\mu q_{\inl}) ~,
\label{zqq}
\ee
where
\be
\delta g = - \frac{C_q}{C_t} \frac{m_t}{8 \pi v} 
\left(\frac{N_{\rm TC}}{3} \right)^{\! 1/2} 
\approx - 2.3 \times 10^{-2}  \frac{C_q}{C_t} 
\left(\frac{N_{\rm TC}}{2} \right)^{\! 1/2} 
~. \label{g-rewrite}
\ee
We will now explore the consequences of this additional coupling.

The first noticeable effect is on the oblique radiative parameter $S$ \cite{sdef},
\be
S \equiv - 16 \pi \left[ {{\rm d} \over {\rm d} q^2} \Pi_{3B}
\right]_{q^2\ =\ 0\ }\ .
\ee
In this model, $S$ is given by
\be
S = S^0 + S^{(t,b)} ~, 
\label{esss}
\ee
where $S^0$ is the technifermion contribution noted earlier in
eq.~(\ref{spar}), and $S^{(t,b)}$ is an additional contribution from the
effective coupling in eq.~(\ref{zqq}) (see ref.~\cite{negs}),
\be
S^{(t,b)} \approx \frac{4}{3 \pi} \delta g 
\ln\left(\frac{\Lambda}{M_Z}\right) < 0 ~, \label{lameql}
\ee
and $\Lambda$ is a scale of order 1 TeV.  This negative contribution to
$S$ is certainly welcome.

Similarly, $\chi^t$ contributes to weak isospin violation, as measured
by the parameter
\be
T \equiv \frac{4}{\alpha v^2} \left[ \Pi_{11}(0) - \Pi_{33}(0) \right] ~,
\ee
where $\Pi_{ii}(q^2)$  are the vacuum
polarizations of the $W^\mu_i$ gauge fields
due to non-standard model physics, with the gauge couplings factored
out.  The operator in eq.~(\ref{operator}) contributes directly
to the $T$ parameter:
\be
T^{(t,b)} \approx - \delta g \frac{3 m_t^2}{\pi^2\alpha v^2}
\ln\left( \frac{\Lambda}{m_t} \right) \approx - 34.0 \ \delta g ~.
\label{direct}
\ee
In addition to the direct isospin violation $T^{(t,b)}$, there are 
``indirect'' contributions to $T$ from the technifermion mass spectrum
which can be only roughly estimated:
\be
T^{0} \sim \frac{N_{\rm TC}}{16 \pi^2 \alpha v^2}
\left( \Sigma_{\up}(0) - \Sigma_{\down}(0) \right)^2  ~,
\ee
where $\Sigma_{\up}(q^2)$ and $\Sigma_{\down}(q^2)$  are the
technifermion self-energies. In this model, the 
indirect isospin violation is due to the last two terms in eq.~(\ref{operator}).
These four-fermion interactions induced by $\chi^t$ exchange
give a one-loop correction to $\Sigma_{\down}$ which is quadratically
divergent:
\be
\Sigma_{\up}(0) - \Sigma_{\down}(0) = - \frac{3}{16 \pi^3} \,
\frac{m_t^2}{v^3} \Lambda^{\prime 2} ~,
\ee
with $\Lambda^{\prime}$ a scale of order 1 TeV, potentially different than 
$\Lambda$.
Putting these contributions together and taking $\Lambda = 1$ TeV in
eq.~(\ref{lameql}), we obtain
\bear
T & \approx & 
4.2 \times 10^{-3} \, \frac{N_{\rm TC}}{2} \,
\left(\frac{\Lambda^\prime}{\rm 1 \, TeV}\right)^{\! 4}
+ 0.79 \frac{C_q}{C_t} \left(\frac{N_{\rm TC}}{2}\right)^{\! 1/2}
~, \label{t-rewrite}
\\ [2mm]
S & \approx & 0.1 N_{TC} - 2.4 \times 10^{-2} 
\frac{C_q}{C_t} \left(\frac{N_{\rm TC}}{2}\right)^{\! 1/2} 
\label{s-rewrite}
\eear
The 2$\sigma$ upper bound $T < 0.71$ from ref.~\cite{negs} then suggests that 
the Yukawa coupling $C_t$ must be at least as large as $C_q$; if 
$\Lambda^\prime$ is not significantly larger than 1 TeV, one finds $C_t \gae C_q$.

In addition to the oblique corrections, the shift in the coupling of the $Z$
boson to third-generation left-handed quarks also makes direct corrections to
several observables measured at the Z-pole: the total $Z$ decay width
$\Gamma_Z$, the peak hadronic cross-section $\sigma_h$, the rate of $Z$
decays to $b$-quarks relative to other hadrons $R_b$, the front-back
asymmetry in $Z$ decays to $b$-quarks $A_{FB}(b)$, and the rates of $Z$
decays to leptons relative to hadrons $R_e,\ R_{\mu},\ R_{\tau}$.  The
oblique and direct corrections that $\chi^t$ causes in the full set of
electroweak observables are summarized in Appendix B.  We derived the
expressions by adapting the analysis of \cite{BuLo} to our model and using
eqs.~(\ref{g-rewrite}), (\ref{t-rewrite}) and (\ref{s-rewrite}) to write the
results in terms of $C_q$ and $C_t$.

We used a least-squares fit to evaluate the models' agreement with the
electroweak data \cite{LEWWG,negs} for different values of $N_{TC}$; the
resulting values of the observables are given in Table \ref{Pred}.  For $N_{TC} = 2$,
a fit setting $\Lambda^\prime = 1$ TeV yields a ratio of Yukawa couplings
\be
{C_q \over C_t} = 0.025 \pm 0.013\ \ .
\ee
The central value has $\chi^2/N_{dof} \approx$ 30.5/21, which
corresponds to a goodness of fit P($N_{dof}$, $\chi^2$) = 8.3\%.  
Leaving both $C_q/C_t$ and $\Lambda^\prime$ free yields best-fit
values
\be
{C_q \over C_t} = 0.017 \pm 0.033 \ \ \ \ \ \ \ \ \ \ \ \ 
\Lambda^\prime = 2.4 \pm 0.36\, {\rm TeV}
\ee
with $\chi^2/N_{dof} \approx $ 28/20 and P($N_{dof}$, $\chi^2$) = 11\%.  The
goodness-of-fit is comparable to, or slightly better than, that of the
standard model ( P($N_{dof}$, $\chi^2$) = 6\% for a Higgs boson mass of 
300 GeV) as evaluated in \cite{negs}.
For the best-fit values of the model parameters, the predicted value of each
observable is within 3$\sigma$ of the experimental value (except for
$A_{LR}$, which is slightly further away).  Moreover, the error ellipses for
the model parameters overlap the region of parameter space ($\Lambda^\prime
\sim 2.3$ TeV, $C_q/C_t < 0.07$) in which all of the observables are within
3$\sigma$ of their experimental values.  In contrast, the fits for $N_{TC} =
4$ are much poorer, with goodness-of-fit less than 1\%, and it is never
possible to have all observables within 3$\sigma$ of their experimental
values.  The larger value of $N_{TC}$ increases $S^0$ enough to prevent
$\Gamma_Z$ and $A_{LR}$ from simultaneously agreeing with experiment.

Overall, a one-doublet technicolor model with an extra technicolored $\chi^t$
scalar gives reasonable agreement with electroweak data only for $N_{TC} =
2$.  The model presented thus far is incomplete, as it does not
provide masses for the light quarks and leptons.  We will see that the
additional physics required to address this goal also causes the vacuum
alignment of $SU(2)_{TC}$ to occur in the pattern that breaks the electroweak
symmetry appropriately.


\begin{table}[htbp]
\begin{center}
\begin{tabular}{|c|l|l|l|l|}\hline\hline
Quantity & Experiment & SM & ${\rm N_{TC} = 2}$ & ${\rm N_{TC} = 4}$
\\\hline \hline
$\Gamma_Z$ & 2.4948 $\pm$ 0.0025 & 2.4966 & 2.4977  & 2.4986 \\
$R_e$ & 20.757 $\pm$ 0.056 & 20.756 & 20.756 & 20.756 \\
$R_\mu$ & 20.783 $\pm$ 0.037 & 20.756 & 20.756 & 20.756  \\
$R_\tau$ & 20.823 $\pm$ 0.050 & 20.756 & 20.756 & 20.756  \\
$\sigma_h$ & 41.486$\pm$ 0.053 & 41.467 & 41.467 & 41.467 \\
$R_b$ & 0.2170 $\pm$ 0.0009 & 0.2158 & 0.2173 & 0.2162  \\
$R_c$ & 0.1734 $\pm$ 0.0008 & 0.1723 & 0.1718 & 0.1720  \\
$A_{FB}^e$ & 0.0160 $\pm$ 0.0024 & 0.0162 & 0.0156  & 0.0151\\
$A_{FB}^\mu$ & 0.0163 $\pm$ 0.0014 & 0.0162 & 0.0156 & 0.0151  \\
$A_{FB}^\tau$ & 0.0192 $\pm$ 0.0018 & 0.0162 & 0.0156 & 0.0151 \\
$A_{\tau}(P_\tau)$ & 0.1411 $\pm$ 0.0064 & 0.1470 & 0.1443 & 0.1424  \\
$A_{e}(P_\tau)$ & 0.1399 $\pm$ 0.0073 & 0.1470 & 0.1443 & 0.1424 \\
$A_{FB}^b$ & 0.0984 $\pm$ 0.0024 & 0.1031 & 0.1014  & 0.1000 \\
$A_{FB}^c$ & 0.0741 $\pm$ 0.0048 & 0.0736 & 0.0722  & 0.0712  \\
$A_{LR}$ & 0.1550 $\pm$ 0.0034 & 0.1470 & 0.1443 & 0.1424  \\
$M_W$ & 80.41 $\pm$ 0.09 & 80.375 & 80.375& 80.375 \\
$g_L^2(\nu N \rightarrow \nu X)$ & 0.3003 $\pm$ 0.0039 & 0.3030 
& 0.3035  & 0.3041 \\
$g_R^2(\nu N \rightarrow \nu X)$ & 0.0323 $\pm$ 0.0033 & 0.0300 
& 0.0302 & 0.0303  \\
$g_{eA}(\nu e \rightarrow \nu e)$ & $-$0.503 $\pm$ 0.018 & $-$0.507 
& $-$0.5076 & $-$0.5083\\
$g_{eV}(\nu e \rightarrow \nu e)$ & $-$0.025 $\pm$ 0.019 & $-$0.037 
& $-$0.036 & $-$0.036\\
$Q_W(Cs)$ & $-$72.11 $\pm$ 0.93 & $-$72.88 & $-$73.04  &  $-$73.20\\
$R_{\mu \tau}$ & 0.9970 $\pm$ 0.0073 & 1.0 & 1.0 & 1.0  \\
\hline\hline
\end{tabular}
\end{center}
\caption{Experimental and standard model values \protect\cite{LEWWG,negs} and
predicted values of electroweak observables for $N_{TC}$ = 2 and 4.
Both $C_q/C_t$ and $\Lambda^\prime$ were set equal to their best-fit
values: for $N_{TC} = 2 (4)$, these are $C_q/C_t$ = 0.017 (0.0027) and 
$\Lambda^\prime$ = 2.4 TeV ( 3.0 TeV).}
\label{Pred}
\end{table}

%
\section{Masses for the Other Quarks and Leptons}
\setcounter{equation}{0}

We turn, now, to addressing the origin of the masses of the other quarks
and leptons.  These much smaller masses can be generated by physics well
above the electroweak scale.  Note that an additional small contribution to 
the top quark's mass may also result from this physics.

One question that will naturally arise when the origins of the other quarks'
masses and mixing angles are considered is the extent to which
flavor-changing neutral currents constrain the model.  Such considerations
are unlikely to place significant limits on the properties of $\chi^t$.  The
strength of the $\chi^t$ state's interactions with the $d$ and $s$ quarks is
not particularly large, being given roughly by the size of the
inter-generational mixing (i.e., of order 0.1-0.01).  Furthermore, $\chi^t$
couples only to left-handed down-type quarks.  So any constraints arising
from $\chi^t$ exchange in the box diagrams for $K^0 \bar K^0$ and $B^0 \bar
B^0$ mixing or the loop diagrams for $b \to s \gamma$ would limit only the
mixing angles of the left-handed quarks.  If the flavor-symmetry-breaking
mixings for down-type quarks are largely in the right-handed sector, extra
FCNC contributions from $\chi^t$ will be suppressed.  Since the $\chi^t$ is
the only new physics that couples to the large mass of the top quark, this
line of reasoning suggests that FCNC need not be a problem in the class of
models we are considering.

\subsection{Weak-doublet technicolor-singlet scalar, $\phi$}

As a simple realization of the higher-scale physics responsible
for the light fermions' masses, we consider the
existence of a scalar, $\phi$, which transforms as $(1,1,2)_{+1}$ under
the (technicolor $\times$ standard model) gauge group. Although its
quantum numbers are the same as those of the standard model Higgs
doublet, the behavior of $\phi$ is considerably different because we
assume that its mass-squared is {\em positive}, as in
ref.~\cite{tcscal}.  Furthermore, $\phi$ need not couple predominantly
to the top quark. Following ref.~\cite{tcscal}, we allow the most
general Yukawa couplings of $\phi$ to technifermions,
\be
\lambda_+ \overline \Psi_{\!\!\inl} \up_{\inr}  i \sigma_2 \phi^\dagger + 
\lambda_- \overline \Psi_{\!\!\inl} \down_{\inr} \phi 
 + {\rm h.c.} ~,
\label{yuk}
\ee
and also to the quarks and leptons.  If $M_{\phi}$ is larger than the
technicolor scale, or if $\lambda_\pm$ is smaller than order one, then
the effect of $\phi$ on the technicolor dynamics is small, as we make
explicit below.  When the
technifermions condense, the interactions in expression 
(\ref{yuk}) give rise to a tadpole term for $\phi$,
\be
\pi f^3 \left(\frac{3}{N_{\rm TC}} \right)^{\! 1/2}
\left( \lambda_+ + \lambda_- \right) 
(1 - \sigma_3) \phi  + {\rm h.c.} ~,
\ee
which leads to a VEV
\be
\langle \phi \rangle = \frac{1}{\sqrt{2}}
\left( \ba{c} 0 \\ f^\prime \ea \right) ~.
\ee
If the quartic scalar operators have small coefficients, then
\be
f^\prime \approx  2 \sqrt{2} \, \pi f^3 \, 
\frac{\lambda_+ + \lambda_-}{M_\phi^2} 
\left(\frac{3}{N_{\rm TC}} \right)^{\! 1/2} ~.
\label{fprime}
\ee
The electroweak symmetry is broken not only by the technifermion condensates,
but also by the VEV of $\phi$, so that the electroweak scale is given by
\be
v = \sqrt{f^2 + f^{\prime 2}} ~.
\label{newvv}
\ee
The four real scalar components of the $\phi$ doublet form an iso-triplet and
an iso-singlet. In general, the iso-triplet mixes with the three
techni-pions, forming the longitudinal $W$ and $Z$, as well as a triplet of
physical pseudo-scalars \cite{carsim}.  Finally, as discussed in Appendix A,
the interaction between $\phi$ and the technifermions affects the vacuum
alignment enough to make even $SU(2)_{TC}$ viable

We assume that the bulk of the top mass is given by the technicolor 
sector, as discussed in section 2. In this case, $f$ cannot be much smaller 
than $v$, suggesting 
\be
f^\prime \ll f \approx v ~,
\label{ineq1}
\ee
so that the longitudinal $W$ and $Z$ are predominantly composed of
techni-pions.

The inclusion of the $\phi$ doublet in the technicolor model with a
weak-doublet techni-scalar provides masses for all the standard fermions
while evading the usual constraints on the standard model Higgs, which
arise from the requirement that the Higgs doublet be responsible for
electroweak symmetry breaking and for the top quark's mass.  The
contributions from $\phi$ to the elements of the quark and lepton mass
matrices are given by $f^\prime/\sqrt{2}$ times the corresponding Yukawa
coupling constants.  The top quark's mass separately receives the
contribution estimated in eq.~(\ref{mtop}).

A lower bound for $f^\prime$ comes from requiring the $b$-quark's Yukawa
coupling constant not to be larger than order one:
\be
f^\prime \gae\sqrt{2} m_b ~.
\label{ineq2}
\ee
Due to the relation~(\ref{fprime}) between the $\phi$ mass and VEV, the 
bounds on $f^\prime$ from eqs.~(\ref{ineq1}) and (\ref{ineq2}) 
 impose constraints on $M_\phi$:
\be
1 \; {\rm TeV} \lae \frac{M_{\phi}}{\sqrt{\lambda_+ + \lambda_-}}
\lae 5 \; {\rm TeV} ~.
\label{mphicon}
\ee

The major advantage of such a model, as far as predicting the fermion 
spectrum is concerned, lies in the fact that the top quark is naturally 
the heaviest.  Note also that it appears that a theory of a composite 
$\phi$, in which the Yukawa couplings are determined, is 
in principle easier to construct than in
the case of the standard model Higgs, because one does not have to worry about 
the $W$, $Z$ and $t$ masses.  However, the low-energy 
Yukawa couplings of $\phi$ are no more 
constrained by theory than those of the standard model Higgs boson.

The presence of heavy weak-doublet scalars 
$\phi$ need not greatly alter the low-energy 
electroweak or flavor-changing neutral current 
phenomenology of the model. First, consider 
the electroweak effects.  Recall (cf. eq.~(\ref{newvv})) 
that the technifermion condensate will generate 
a small vev ($f'$) for $\phi$, and it is the 
combination of decay constants $f^2 + f'^2$ 
which now equals $v \approx 246$ GeV.  The factor 
$v^2$ in eq.~(\ref{vchg})  therefore 
becomes an $f^2$ and the expression 
(\ref{g-rewrite}) for the coupling shift will 
be multiplied by the ratio $f^2/v^2$.  
Yet the net effect must be small in order 
for our analysis to remain self-consistent: 
according to eq.~(\ref{mtop}), keeping 
the top quark mass fixed while lowering $f$ requires either raising 
the values of the Yukawa couplings $C_q$ and $C_t$ or reducing 
the mass of $\chi_t$ -- both of which are problematic.  
A further effect of the presence of $\phi$ is to make an additional 
contribution to the $S$ parameter (\ref{esss}); as long as $f' \ll v$, 
however, this contribution will be negligible \cite{tcscal}.  
There are also contributions from $\phi$ to the $T$ parameter to the 
extent that the coupling of technifermions to $\phi$ violates weak 
isospin; again, these can be made small \cite{tcscal}.  Finally, 
we come to flavor-changing neutral currents.  The size of the 
contributions $\phi$ makes to $K^0 \bar K^0$ mixing, $B^0 \bar B^0$ 
mixing, or $b\to s \gamma$ is proportional to powers of 
the $\phi$ state's Yukawa couplings to quarks.  Such contributions 
have been found \cite{tcscal,kagan2,massless} to be significant in the case 
where the Yukawa coupling of $\phi$ to $t$ is large enough to 
generate the full top quark mass.  In our model, however, 
$\phi$ need contribute no more to $m_t$ than to $m_b$; this 
suppresses the extra FCNC contributions by several powers of $m_b/m_t$,
making them far less restrictive.

\subsection{Weak-doublet technicolored scalars, $\chi_f$}

The situation is different in models with additional weak-doublet
technicolored scalars.  In addition to $\chi^t$, there are only three
scalar representations of the (technicolor $\times$ SM) gauge group,
$\chi_b$, $\chi_\tau$ and $\chi_{\nu_\tau}$, that can have
Yukawa couplings involving a technifermion and a standard fermion:
\be
\chi_b: (\overline{N_{\rm TC}}, 3, 2)_{-2/3} \;\: , 
\;\; \chi_\tau: (\overline{N_{\rm TC}}, 1, 2)_{-2} \;\; , 
\;\; \chi_{\nu_\tau}: (\overline{N_{\rm TC}}, 1, 2)_{0}  ~.
\ee

The most general Yukawa couplings of $\chi_b$, for conveniently
chosen quark eigenstates, are given by
\be 
{\cal L}_{b} = C_q^b\, \overline{q^3_{\inl}}\, \up_{\inr}\, \chi_b
+  C_q^{b\prime}\, \overline{q^2_{\inl}}\, \up_{\inr}\, \chi_b
+ C_b\, \overline \Psi_{\!\!\inl}\, b_{\inr}\, i \sigma_2\,\chi_b^\dagger 
+ {\rm h.c.}
\ee
Note that after a $U(3)$ flavor redefinition as in the case of
eq.~(\ref{tyuk}), only one down-type right-handed quark, namely $b_R$,
couples to $\chi_b$.  For left-handed quarks, however, only a $U(2)$
flavor transformation is available, because $q_{\inl}^3$ is already
defined by eq.~(\ref{tyuk}).  Therefore, $\chi_b$ couples to both
$q_{\inl}^2$ and $q_{\inl}^3$.  As a result, both a $b$ quark mass,
\be
m_b \approx \frac{C_q^b C_b}{M_{\chi_b}^2}\pi f^3 
\left(\frac{3}{N_{\rm TC}} \right)^{\! 1/2} ~,
\label{mbottom}
\ee
and a $b-s$ quark mixing are induced (it is, thus, necessary that 
$C_q^b \gg C_q^{b\prime}$). 
Comparing eqs.~(\ref{mtop}) and (\ref{mbottom}) one can see that the ratio 
$m_t/m_b \approx 40$ can have its origin in a scalar mass ratio
\be
\frac{M_{\chi_b}}{M_{\chi^t}} \approx 6.5 ~,
\ee
instead of a large ratio of Yukawa coupling constants.

Likewise, the most general Yukawa couplings of $\chi_\tau$ 
are given by
\be 
{\cal L}_{\tau} = C_l \overline{l^3_{\inl}} \up_{\inr} \chi_\tau
+ C_\tau \overline \Psi_{\!\!\inl} \tau_{\inr} i \sigma_2\chi_\tau^\dagger 
+ {\rm h.c.} ~,
\ee
where $l^3_{\inl} = (\nu_\tau, \tau)^{\top}$ is the left-handed third
generation lepton.  The $\tau$ mass is produced by the exchange of
$\chi_\tau$, with the condition
\be
M_{\chi_\tau} \approx 5.7 \; {\rm TeV} \, \sqrt{C_l C_\tau}
\left(\frac{2}{N_{\rm TC}} \right)^{\! 1/4} ~.
\ee
Including the scalar $\chi_{\nu_\tau}$ would be useful only for
producing a Dirac mass for $\nu_\tau$.

A second generation of weak-doublet techni-scalars would give masses to
the second generation of quarks and leptons (this scenario is discussed
briefly for the case of weak-singlet techni-scalars in \cite{kagan2}).
It is possible then to trade all the large ratios of Yukawa couplings
required in the standard model for smaller ratios of scalar masses, with the 
hope that the scalar spectrum is correctly produced by the high energy theory
responsible for scalar compositness or supersymmetry breaking.

Exchange of the numerous technicolored scalars $\chi^f$ would make 
contributions to electroweak radiative corrections analogous to those 
from $\chi^t$.  However, such effects are suppressed relative 
to the effects of $\chi^t$ by the ratio of the lighter fermion mass 
to $m_t$.  Since the minimum suppression is by a factor of 40, 
these corrections are small enough to ignore.

\section{Composite Scalars}
\setcounter{equation}{0}

In the previous sections we showed that the inclusion of scalar fields
in a minimal technicolor model may be useful in explaining certain features
of the quark and lepton spectrum, such as the heaviness of the top quark,
or the mass hierarchy between the third generation and the others.
However, in the absence of supersymmetry, the existence of scalar fields 
much lighter than the Planck scale is natural only if these scalars are 
composite. In this section we discuss various possibilities for the 
existence of fermion-antifermion states bound by new
dynamics at a scale of order a TeV, or higher.

\subsection{Quark-technifermion bound states}

The simplest possibility that leads to compositness for the scalars discussed 
in sections 2 and 4, is the existence of a new non-confining gauge interaction 
that binds together standard fermions and technifermions.
In this case, the $\chi^t$ techniscalar that is responsible for the top 
quark mass can be a $\overline{t}_R \Psi_L$ or a $\overline{N}_R q_L^3$ state.
More generally, both these states are present, and because they have the same
transformation properties under the (technicolor $\times$ SM) gauge group,
a large mixing between them is induced by technicolor interactions. The
$\overline{t}_R \Psi_L$ composite scalar (labeled $\chi^t_R$)
has a large Yukawa coupling to the
$t_R$ and $\overline{\Psi}_L$ fields, while the $\overline{N}_R q_L^3$ 
composite scalar ($\chi^t_L$) couples to $\overline{q}_L^3$ and $N_R$. 
This situation is reminiscent of the 
supersymmetric technicolor models of refs.~\cite{kagan1,susytc}, 
where a  combination
of superpotential holomorphy and gauge anomaly cancellation requires 
the existence of {\it two} techniscalars which mix. 
Due to the scalar mixing, the exchange of the two physical scalar states 
gives rise to four-fermion operators as in eq.~(\ref{operator}), but with 
modified coefficients: $M_{\chi^t}^2$ is replaced  by
\[
M_{\chi^t_L}^2 - \frac{M_{\chi^t_{LR}}^4}{M_{\chi^t_R}^2} \;\; , \;\;
M_{\chi^t_R}^2 - \frac{M_{\chi^t_{LR}}^4}{M_{\chi^t_L}^2} \;\; , \;\;
\frac{M_{\chi^t_L}^2 M_{\chi^t_R}^2}{M_{\chi^t_{LR}}^2} - M_{\chi^t_{LR}}^2
\]
respectively in the first, second, and last two terms of ${\cal L}_{\rm 4F}$
[also in eqs.~(\ref{mtop}) and (\ref{mchival})], where $M_{\chi^t_L}$ and
$M_{\chi^t_R}$ are the $\chi^t_L$ and $\chi^t_R$ masses, and $M_{\chi^t_{LR}}$
is the mass mixing.
Consequently, the results of section 4 survive, modified only by having the
ratio $C_q/C_t$ multiplied by $M_{\chi^t_R}^2/M_{\chi^t_{LR}}^2$.

An example of the non-confining interaction that can bind together the top
quark fields and technifermions is a $U(1)_{\rm new}$ gauge symmetry,
attractive in the $\overline{t}_R \Psi_L$ and $\overline{N}_R q_L^3$
channels, and broken at a scale in the TeV range.  In order to form
sufficiently narrow bound states, this interaction has to be rather strong,
though it need not be strong enough to produce fermion/anti-fermion
condensates.  Avoiding a Landau pole for the $U(1)_{\rm new}$ gauge coupling
requires the embedding of $U(1)_{\rm new}$ in a non-Abelian gauge group at a
scale just slightly higher than the composite scalar masses.  There are also
several constraints on the $U(1)_{\rm new}$ charges. First, exchange of the
heavy $U(1)_{\rm new}$ gauge boson gives rise to four-technifermion
operators; to prevent these operators from making a large contribution to the
isospin breaking parameter $T$, the $U(1)_{\rm new}$ charges of $P_R$ and
$N_R$ must be equal.  Second, requiring anomaly cancellation imposes
relations between the various fermions' $U(1)_{new}$ charges.  These
relations constrain the coefficients of the four-fermion operators induced by
the strong $U(1)_{new}$. For example, in the simplest scenario the only
fields charged under $U(1)_{\rm new}$ are the third generation fermions
(including the right-handed neutrino\footnote{Otherwise the
  anomaly-cancellation conditions would make the $\overline{t}_R \Psi_L$ or  $\overline{N}_R q_L^3$ channels repulsive.}  ) and the technifermions (with
$P_R$ and $N_R$ having equal charges).  In this case, the $\overline{\tau}_R
l_L^3$ and $\overline{\nu}_{\tau_R} l_L^3$ channels turn out to be much more
attractive than $\overline{t}_R \Psi_L$ and $\overline{N}_R q_L^3$. As a
result, a couple of composite $\phi$ scalars will form and will even be
narrower than the $\chi^t_{L,R}$ scalars.  Other attractive channels lead to
the formation of leptoquarks and color-octet scalars.  The current limited
knowledge of strongly coupled field theory does not allow us to decide
whether the $U(1)_{\rm new}$ gauge group gives rise to the precise scalar
spectrum we need.

An alternative method for producing top-technifermion bound states from
non-confining interactions might arise in composite models.  If the top quark
and technifermion fields were composites created by some underlying
high-energy theory, additional top-technifermion interactions able to produce
the requisite scalar states could be present.


\subsection{New fermions as constituents}

An alternative possibility is that the 
 composite $\chi^t$ scalar could be produced by the action of a new
strong gauge group $SU(n)_{NJL}$ on a set of additional fermions charged under 
that
group. 

 For example, consider two Dirac fermions, $A$ and $B$, which
transform under the gauge groups as shown in Table 2.
\begin{table}[htbp]
\centering
\begin{tabular}{|c||c|c|c|c|c|}\hline
 & $SU(n)_{NJL}$ & $SU(4)_{\rm TC}$ & $SU(3)_C$ & $SU(2)_W$ & $U(1)_{Y}$ \vbr
\\\hline \hline
$A$ & \Yfund & \Yfund & {\bf 1} & {\bf 1} & y \vbr\\  \hline
$B$ & \Yfund & {\bf 1} & \Yfund & \Yfund & 4/3 + y \vbr \\ \hline
\end{tabular}
\label{Sp}
\parbox{3in}{\caption{Fermion constituents of $\chi^t$.}}
\end{table}
Note that these new fermions are vectorlike, so that they do not have large
contributions to the $S$ and $T$ parameters.  If the $SU(n)_{NJL}$ gauge
group is spontaneously broken and under-critical, a $\overline{B} A$ scalar
with positive mass-squared will be formed, as can be proven in the large $n$
limit (as has been shown \cite{tmode} in the Nambu--Jona-Lasinio model
\cite{njl}). This bound state has the transformation properties of $\chi^t$.
Inducing the Yukawa interactions of $\chi^t$ requires additional 4-fermion
operators:
\be 
{\cal L}_{\rm 4f} = \frac{1}{M^2} \overline q_{\inl}^3 \down_{\inr} 
\overline{B} A
+ \frac{1}{M^{\prime 2}}
\epsilon_{ij} \overline \Psi_{\!\!\inl_i} t_{\inr} \overline{A}_j B
 + {\rm h.c.}~ ~.
\ee
which must be provided by physics at higher scales.
The $\chi^f$ and $\phi$ scalars may have a similar origin.

\subsection{Strongly coupled ETC}

Finally, we note that a composite technicolor-singlet scalar $\phi$ state and
its couplings to fermions could result from strongly-coupled ETC interactions
between standard fermions and technifermions \cite{strongETC}. The fact that
$\phi$ need not provide the entire large mass of the top quark would provide
useful flexibility in keeping the model's phenomenology consistent with
experiment.  Given that the quarks and technifermions need to belong to the
same ETC multiplets, it is even possible that strongly-coupled ETC could give
rise to a composite $\chi^t$ techniscalar.

\section{Conclusions}

We have introduced a class of technicolor models in which the top quark mass
is produced by exchange of a weak-doublet technicolored scalar multiplet.
Such models can explain the large top quark mass while remaining consistent
with precision electroweak data.

A single scalar multiplet $\chi^t$ with $SU(N)_{TC} \times SU(3)_C \times
SU(2)_W \times U(1)$ charges $(N_{TC}, 3, 2)_{4/3}$ gives mass only to a
single up-type quark.  The hypercharge quantum number prevents this scalar
from coupling to the right-handed bottom quark and generating $m_b$. Thus, a
model containing a $\chi^t$ scalar naturally produces both the large top
quark mass and the large splitting between $m_t$ and $m_b$.  We have
identified several promising dynamical mechanisms through which $\chi^t$
could be created as a fermion/anti-fermion bound state.  These include models
with no new fermions and a strong $U(1)$ gauge interactions, models with new
fermions as sub-constituents of $\chi_t$, and models with strongly-coupled
extended technicolor interactions.

A minimal model including one-doublet $SU(2)_{TC}$ technicolor and a $\chi^t$
scalar is in agreement with the precision electroweak data.  For
larger $N_{TC}$ the agreement is poor; additional physics would be required
to ammend the corrections to the electroweak observables.

The simplest way of generating the masses of the lighter fermions is to
include a weak-doublet technicolor-singlet scalar $\phi$.  Including $\phi$
has several virtues.  The vacuum alignment of $SU(2)_{TC}$ produces the
correct electroweak symmetry breaking pattern in the presence of a $\phi$
scalar.  The presence of $\phi$ need not modify the successful match between
the minimal model and the electroweak data.  The $\phi$ scalar can be readily
generated by the same dynamics that produces a $\chi^t$ bound state.  So we
have a complete and appealing package.

Another alternative is for the masses of some of the lighter fermions to be
created by additional weak-doublet technicolored scalars $\chi_f$.  This
could neatly explain fermion mass hierarchies in terms of relationships among
the $\chi_f$ masses while leaving intact the agreement between the predicted
and measured electroweak observables.

\vspace{12pt} \centerline{\bf Acknowledgments} \vspace{2mm}

We thank Sekhar Chivukula, Alex Kagan, and John Terning for useful
discussions.  E.H.S. thanks the Theoretical Physics Group at Fermilab for its
hospitality during the course of this work; both authors are grateful to the
Aspen Center for Physics for its hospitality during this work's completion.
E.H.S. acknowledges the support of the Faculty Early Career Development
(CAREER) program and the DOE Outstanding Junior Investigator program. {\em
  This work was supported in part by the National Science Foundation under
  grant PHY-9501249, and by the Department of Energy under grant
  DE-FG02-91ER40676.}

\section*{Appendix A: Vacuum Alignment in $SU(2)$ Technicolor}
\renewcommand{\theequation}{A.\arabic{equation}}
\setcounter{equation}{0}

In this Appendix we show that $SU(2)_{\rm TC}$ technicolor correctly breaks
the electroweak symmetry in the presence of the weak-doublet $\phi$ scalar
discussed in section 4.1.

We start by reviewing the argument against minimal $SU(2)_{\rm TC}$
technicolor \cite{align,farhi}. The $SU(2)_{\rm TC}$ group has only real 
representations, so that, in the absence of the electroweak 
interactions, there is an $SU(4)_F$ chiral symmetry acting on the 
$P_L, N_L, P^c, N^c$ technifermions ($P^c$ and $N^c$ are the charge conjugate 
fermions corresponding to $P_R$ and $N_R$). 
At the scale where $SU(2)_{\rm TC}$ becomes strong, the technifermions 
condense and break the chiral symmetry down to $Sp(4)$. 
Therefore, the vacuum manifold is $Sp(4)$ symmetric, which implies that
the two condensation channels,
$\langle P_L P^c \rangle = \langle N_L N^c \rangle$ and 
$\langle P_L N_L \rangle = \langle P^c N^c \rangle$,  
are equally attractive.

In the presence of the $SU(2)_W \times U(1)_Y$ interactions, the degeneracy
of the vacuum manifold is lifted, Exchange of the $SU(2)_W\times U(1)_Y$
gauge bosons, $W_\mu^a$ and $B_\mu$, contributes at one loop to the $P_L N_L$
and $P^c N^c$ dynamical masses, respectively, but makes no contribution to
the $P_L P^c$ or $N_L N^c$ dynamical masses.  The net effect is 
\be
\delta\left( M_{P_L N_L} + M_{P^c N^c} \right) \approx
\frac{3 g^2 + g^{\prime 2}}{ 8 \pi^2} M_{P_L N_L}
\ln\left(\frac{M_{\rm TC} }{M_{P_L N_L}}\right) > 0 ~,
\label{wrong}
\ee
where $M_{P_L N_L} \sim 1$ TeV (given by scaling the constituent quark mass
from QCD), $M_{\rm TC}$ is a physical cut-off of order $4\pi M_{P_L N_L}$,
while $g$ and $g^\prime$ are the $SU(2)_W \times U(1)_Y$ gauge couplings.
The increase in the dynamical masses implies that the $SU(2)_W \times U(1)_Y$
interactions make the $\langle P_L N_L \rangle \approx \langle P^c N^c
\rangle$ channel more attractive than $\langle P_L P^c \rangle = \langle N_L
N^c \rangle$, leading to a complete breaking of the electroweak symmetry.
Thus, minimal $SU(2)_{\rm TC}$ technicolor is not viable on its own.

However, the vacuum is tilted in the wrong direction only by the electroweak
interactions, which are a small perturbation on the technicolor dynamics.
Additional interactions of the technifermions may easily change
the vacuum alignment. Consider the effect of the $\phi$ scalar that
communicates electroweak symmetry breaking to the lighter quarks and leptons in the scenarios discussed in section 4.1.
In the vacuum where $\langle P_L P^c \rangle = \langle N_L N^c \rangle$,
$\phi$ acquires a vev which enhances the $P_L P^c$ and $N_L N^c$ masses by an
amount [see eqs.~(\ref{yuk}) and (\ref{fprime})]
\be
\delta\left( M_{P_L P^c} + M_{N_L N^c} \right) \approx
\frac{(\lambda_+ + \lambda_-)^2}{M_\phi^2} 2 \pi f^3 
\left(\frac{3}{N_{\rm TC}} \right)^{\! 1/2} > 0 ~.
\label{good}
\ee
The correct symmetry breaking pattern, $SU(2)_W \times U(1)_Y \rightarrow
U(1)_Q$, requires $\langle P_L P^c \rangle = \langle N_L N^c \rangle$ to be
the most attractive channel, which is satisfied provided the contribution in
eq.~(\ref{good}) is larger than the one in eq.~(\ref{wrong}).  This is
equivalent to the requirement
\be
\frac{M_{\phi}}{\lambda_+ + \lambda_-}
\lae 1.6 \; {\rm TeV} ~,
\ee
which is consistent with eq.~(\ref{mphicon}) and the assumption that 
the $\lambda_\pm$ Yukawa couplings are of order one.

\section*{Appendix B: Electroweak Observables}
\renewcommand{\theequation}{B.\arabic{equation}}
\setcounter{equation}{0}

In this Appendix, adapting the analysis of \cite{BuLo}, 
we present the expressions for the electroweak observables in our models 
relative to those in the standard model. 
The contributions from a minimal technicolor sector and $\chi^t$
are included explicitly; as argued in the text, the effects 
of additional $\phi$ or $\chi^f$ scalars would be negligible
by comparison. 

Using the convenient notation
\be
r \equiv 10^{-2} \sqrt{\frac{N_{\rm TC}}{2}} ~,
\ee
the values of the electroweak observables may be expressed as follows

\[
 \Gamma_Z = \left( \Gamma_Z \right)_{\rm SM} \left[1 -  
 7.7\, r^2 - 0.44\, r^2 
\left({\Lambda^{\prime}\over 1{\rm TeV}}\right)^{\! 4}
+ 2.5 \, r {C_q\over{C_t}}\right]
\nonumber 
\]
\[
 R_{e,\mu,\tau}= \left( R_{e,\mu,\tau} \right)_{\rm SM} \left[1 -  
5.8\, r^2 - 
8.2 \times 10^{-2}\, r^2 
  \left({\Lambda^{\prime}\over 1{\rm TeV}}\right)^{\! 4}+ 
2.5 \, r {C_q\over{C_t}} \right]
\nonumber
\]
\[
 \sigma_h = \left( \sigma_h \right)_{\rm SM} \left[1 +  
  0.44\, r^2  + 6.7 \times 10^{-3}\, r^2 \left(
{\Lambda^{\prime}\over 1{\rm TeV}}\right)^{\! 4}- 
0.95 \, r {C_q\over{C_t}}\right]
\nonumber
\]
\[
 R_b = \left( R_b \right)_{\rm SM} \left[ 1 + 
1.3\, r^2 - 1.7\times 10^{-2}\, r^2 \left(
{\Lambda^{\prime}\over 1{\rm TeV}} \right)^{\! 4}+ 
8.3 \, r {C_q\over{C_t}}\right]
\nonumber
\]
\[
 R_c = \left( R_c \right)_{\rm SM} \left[1  +  
2.6 \, r^2 - 4.2 \times 10^{-2} \, r^2 
\left( {\Lambda^{\prime}\over 1{\rm TeV}}\right)^{\! 4}- 
2.3 \, r {C_q\over{C_t}}\right]
\nonumber
\]
\[
 A_{FB}^{e,\mu,\tau} = \left( A_{FB}^{e,\mu,\tau} \right)_{\rm SM}  
 - 14 \, r^2 - 0.2 \, r^2 \left(
   {\Lambda^{\prime}\over 1{\rm TeV}} \right)^{\! 4}+ 
0.39 \, r {C_q\over{C_t}}
\nonumber
\]
\[
 A_{e,\tau}(P_\tau) = \left( A_{e,\tau}(P_\tau) \right)_{\rm SM} 
 - 57 \, r^2 - 0.84 \, r^2 \left(
{\Lambda^{\prime}\over 1{\rm TeV}} \right)^{\! 4}+ 
1.6 \, r {C_q\over{C_t}} 
\nonumber
\]
\[
 A_{FB}^b = \left( A_{FB}^b \right)_{\rm SM} 
 -  380\, r^2 - 0.56\, r^2 \left(
{\Lambda^{\prime}\over 1{\rm TeV}} \right)^{\! 4}+ 
1.2 \, r {C_q\over{C_t}}
\nonumber
\]
\[
 A_{FB}^c = \left( A_{FB}^c \right)_{\rm SM} 
 -  290\, r^2 - 0.43 \, r^2 \left(
{\Lambda^{\prime}\over 1{\rm TeV}} \right)^{\! 4}+ 
0.85 \, r {C_q\over{C_t}}
\nonumber
\]
\[
 A_{LR} = \left( A_{LR} \right)_{\rm SM} 
 -  57\, r^2 - 0.84\, r^2 \left(
{\Lambda^{\prime}\over 1{\rm TeV}}\right)^{\! 4}+ 
5.6 \, r {C_q\over{C_t}}
\nonumber
\]
\[
 M_W = \left( M_W \right)_{\rm SM} \left[ 1 - 
 7.2 \, r^2  -  0.23 \, r^2 \left(
{\Lambda^{\prime}\over 1{\rm TeV}}\right)^{\! 4}+ 
0.44 \, r  {C_q\over{C_t}}\right]
\nonumber
\]
\[
 g_L^2(\nu N \rightarrow \nu X) =
\left( g_L^2(\nu N \rightarrow \nu X) \right)_{\rm SM} 
 - 5.4 \, r^2 - 0.28 \, r^2 \left(
{\Lambda^{\prime}\over 1{\rm TeV}}\right)^{\! 4}+ 
0.53 \, r {C_q\over{C_t}}
\nonumber
\]
\[
 g_R^2(\nu N \rightarrow \nu X) =
\left( g_R^2(\nu N \rightarrow \nu X) \right)_{\rm SM}  
 +  1.9 \, r^2 - 8.0 \times 10^{-3}\, r^2 \left(
{\Lambda^{\prime}\over 1{\rm TeV}}\right)^{\! 4}- 
0.017 \, r {C_q\over{C_t}}
\nonumber
\]
\[
 g_{eA}(\nu e \rightarrow \nu e) =
\left( g_{eA}(\nu e \rightarrow \nu e) \right)_{\rm SM}  
 - 0.17 \, r^2 \left(
{\Lambda^{\prime}\over 1{\rm TeV}} \right)^{\! 4}- 
0.31 \, r {C_q\over{C_t}}
\nonumber
\]
\[
 g_{eV}(\nu e \rightarrow \nu e) =
\left( g_{eV}(\nu e \rightarrow \nu e) \right)_{\rm SM} 
 + 15 \, r^2  - 0.23 \, r^2 \left(
{\Lambda^{\prime}\over 1{\rm TeV}}\right)^{\! 4}- 
0.44 \, r {C_q\over{C_t}}
\nonumber
\]
\[
 Q_W(Cs) = \left( Q_W(Cs) \right)_{\rm SM} 
 -  1.6 \times 10^3 \, r^2 + 0.47 \, r^2 \left(
{\Lambda^{\prime}\over 1{\rm TeV}}\right)^{\! 4}+ 
1.0 \, r {C_q\over{C_t}}
\nonumber
\]
\be
R_{\mu \tau} \equiv {\Gamma(\tau \to \mu \nu \bar\nu)\over
\Gamma(\mu\to e \nu \bar \nu)} = R_{\mu \tau}^{\rm SM} ~.
\ee

\newcommand{\np}{{\it Nucl.\ Phys.}\ {\bf B}}
\newcommand{\pr}{{\it Phys.\ Rev.}\ }
\newcommand{\prd}{{\it Phys.\ Rev.}\ {\bf D}}
\newcommand{\prp}{{\it Phys.\ Rep.}\ }
\newcommand{\prl}{{\it Phys.\ Rev.\ Lett.}\ }
\newcommand{\pl}{{\it Phys.\ Lett.}\ {\bf B}}
\newcommand{\ptp}{{\it Prog.\ Theor.\ Phys.}\ }
\newcommand{\ap}{{\it Ann.\ Phys.}\ }
\newcommand{\intl}{{\it Int.\ J.\ Mod.\ Phys.}\ {\bf A}}

\vfil 

\begin{thebibliography}{99}
\frenchspacing

\bibitem{tmode} Y.~Nambu, Chicago preprints EFI 88-39 (1988), EFI 88-62
  (1988), and EFI 89-08 (1989); V.A.~Miransky, M.~Tanabashi, and K.~Yamawaki,
  \pl {221} (1989) 177 and {\it Mod. Phys. Lett.} {\bf 4} (1989) 1043;
  W.J.~Marciano, \prl {\bf 62} (1989) 2793 and \prd {\bf 41} (1990) 219;
  W.A.~Bardeen, C.T.~Hill and M.~Lindner, \prd {\bf 41} (1990) 1647;
  C.T.~Hill, \pl {\bf 266} (1991) 419.
  
\bibitem{tmoderec} B.A.~Dobrescu and C.T.~Hill, FERMILAB-PUB-97/409-T,
  EFI-97-55, Dec 1997, hep-ph/9712319; \
M.~Lindner and G.~Triantaphyllou, TUM-HEP-308-98, Mar 1998,
hep-ph/9803383. 

\bibitem{etc} S.~Dimopoulos and L.~Susskind, \np{\bf 155} (1979) 237;
  E.~Eichten and K.~Lane, \pl {\bf 90} (1980) 125.

\bibitem{topassist} C.T.~Hill, \pl {\bf 345} (1995) 483, hep-ph/9411426
  K.~Lane and E.~Eichten, \pl {\bf 352} (1995) 382, hep-ph/9503433;
  G.~Buchalla, G.~Burdman, C.T.~Hill and D.~Kominis, \prd {\bf 53} (1996)
  5185, hep-ph/9510376; K.D.~Lane, \prd {\bf 54} (1996) 2204, hep-ph/9602221;
  K.D.~Lane, hep-ph/9805254, to appear in Phys. Lett. B; M.B.~Popovic and
  E.H.~Simmons, hep-ph/9806287, to appear in Phys. Rev. D.

\bibitem{ncETC} R.S.~Chivukula, E.H.~Simmons, J.~Terning, \pl {\bf 331}
  (1994) 383, hep-ph/9404209; R.S.~Chivukula, E.H.~Simmons, J.~Terning,
  \prd {\bf 53} (1996) 5258, hep-ph/9506427; E.H.~Simmons, \prd {\bf 55}
  (1997) 5494, hep-ph/9612402.

\bibitem{tcscal} E.~H.~Simmons, \np {\bf 312} (1989) 253.

\bibitem{carsim} C.~D.~Carone and E.~H.~Simmons, \np {\bf 397} (1993)
    591.

\bibitem{massless} C.~D.~Carone and H.~Georgi, \prd {\bf 49} (1993) 1427, 
hep-ph/9308205.

\bibitem{phen} C.~D.~Carone, E.~H.~Simmons and Y.~Su, \pl {\bf 344}
  (1995) 287; \ C.~D.~Carone and M.~Golden, 1994.

\bibitem{kagan1} A.~Kagan, {\it Proceedings of the 15th Johns Hopkins
    Workshop on Current Problems in Particle Theory}, G.~Domokos and
  S.~Kovesi-Domokos eds. (World Scientific, Singapore, 1992), p.217.

\bibitem{kagan2} A.~Kagan, {\em Phys. Rev.} {\bf D51} (1995) 6196,
  hep-ph/9409215.

\bibitem{susytc} B.~A.~Dobrescu, {\em Nucl. Phys.} {\bf B449} (1995)
  462, hep-ph/9504399.

\bibitem{negs} B.~A.~Dobrescu and J.~Terning, Phys.~Lett.~{\bf B 416}
  (1998) 129, hep-ph/9709297.

\bibitem{tcsd} S.~Samuel, \np {\bf 347} (1990) 625.

\bibitem{bostc} M.~Dine, A.~Kagan and S.~Samuel, \pl {\bf 243} (1990)
  250;\\ A.~Kagan and S.~Samuel, \pl {\bf 270} (1991) 37;\ \pl {\bf 252}
  (1990) 605;\ \intl {\bf 7} (1992) 1123.

\bibitem{strongETC} R.~S.~Chivukula, A.~G.~Cohen and K.~Lane, \np {\bf
    343} (1990) 554; \ T.~Appelquist, J.~Terning and
  L.~C.~R.~Wijewardhana, \prd {\bf 44} (1991) 871.
  
\bibitem{lykken} J.D.~Lykken, \prd {\bf 54} (1996) 3693, hep-th/9603133;
  N.~Arkani-Hamed, S.~Dimopoulos, and G.~Dvali. SLAC-PUB-7769. Submitted to
  \pl , hep-ph/9803315.

\bibitem{peskin} M. Golden and L. Randall, \np {\bf 361} (1991) 3; B.
  Holdom and J. Terning, \pl {\bf 247} (1990) 88; M.~E.~Peskin and
  T.~Takeuchi, \prd {\bf 46} (1992) 381; W.
  J.~Marciano and J. L.~Rosner, \prl {\bf 65} (1990) 2963 [Erratum ibid.
  {\bf 68} (1992) 898].

\bibitem{sdef} M.~E.~Peskin and T.~Takeuchi, \prl {\bf 65} (1990) 964

\bibitem{align} M. Peskin, {\em Nucl. Phys.} {\bf B175} (1980) 197; J.
  Preskill, {\em Nucl. Phys.} {\bf B177} (1981) 21.

\bibitem{BuLo} C.P.~Burgess, S.~Godfrey, H.~Koenig, D.~London, and
  I.~Maksymyk, \prd {\bf 49} (1994) 6115, hep-ph/9312291.

\bibitem{LEWWG} D. Abbaneo et al. (SLD Heavy Flavor Group and OPAL 
Collaboration and L3 Collaboration and DELPHI Collaboration and ALEPH 
Collaboration and LEP Electroweak Working Group), `A Combination of
Preliminary Electroweak Measurements and Constraints on the Standard Model,'
CERN-PPE-97-154. December 1997.
 
\bibitem{njl} Y.~Nambu and G.~Jona-Lasinio, Phys. Rev. {\bf 122} (1961) 345;
Y.~Nambu and G.~Jona-Lasinio, Phys. Rev. {\bf 124} (1961) 246.

\bibitem{farhi} E.~Farhi and L.~Susskind, Phys. Rept. {\bf74 } (1981) 277.

\end{thebibliography}
\end{document}